\documentclass[fleqn,usenatbib,onecolumn]{mnras}

\usepackage{newtxtext,newtxmath}
\usepackage{orcidlink}
\usepackage[T1]{fontenc}

\DeclareRobustCommand{\VAN}[3]{#2}
\let\VANthebibliography\thebibliography
\def\thebibliography{\DeclareRobustCommand{\VAN}[3]{##3}\VANthebibliography}

\newcommand{\rmd}{{\rm d}}
\newcommand{\rme}{{\rm e}}
\newcommand{\rmi}{{\rm i}}

\usepackage{graphicx}	% Including figure files
\usepackage{amsmath}	% Advanced maths commands

\title{Pulsar scintillation arcs formed from correlated volume scattering}

\author[T. Kramer]{
Tobias Kramer,$^{1,2}$\orcidlink{0000-0003-1106-3587}
\\
$^{1}$Institute for Theoretical Physics, Johannes Kepler University Linz, Altenberger Str.\ 69, 4040 Linz, Austria\\
$^{2}$Department of Physics, Harvard University, 17 Oxford Street, Cambridge, MA 02138, USA\\
}

\date{Accepted XXX. Received YYY; in original form ZZZ}

\pubyear{\the\year{}}

\begin{document}
\label{firstpage}
\pagerange{\pageref{firstpage}--\pageref{lastpage}}
\maketitle

\begin{abstract}
Radio waves propagating through the interstellar medium are influenced by variations in plasma density. 
For spatially localised plasma structures along the line of sight, time-delay Doppler analyses of pulsars often reveal scintillation arcs in the secondary spectrum, frequently exhibiting a parabolic morphology.
In the thin-screen approximation, the arc curvature is commonly used to infer the distance to the plasma concentration, which is modelled --via Kirchhoff–Fresnel diffraction theory-- as an effective phase screen imposed by the column density of a localised disturbance.
Here, we identify several limitations of this model that necessitate a fully three-dimensional treatment, without reducing the problem to a projected screen density. 
When volume propagation is considered, asymmetries and a richer variety of features emerge in the secondary spectrum compared to those predicted by the thin-screen approximation. 
For some cases the arc curvature varies depending on the three-dimensional structure of the plasma, rendering it a less reliable indicator of distance.
We conjecture that these phenomena are linked to the onset of branched flow produced by a sequence of weak but correlated scattering events.
\end{abstract}


%%%%%%%%%%%%%%%%%%%%%%%%%%%%%%%%%%%%%%%%%%%%%%%%%%

%%%%%%%%%%%%%%%%% BODY OF PAPER %%%%%%%%%%%%%%%%%%

\section{Introduction} \label{sec:intro}

Pulsar scintillation arcs arise from the scattering of radio signals with plasma concentrations in the interstellar medium (ISM) (\cite{Stinebring2001}).
The curvature of the scintillation arcs serves as an important proxy to determine the distance of the discrete scattering structures to the observer (\cite{Stinebring2022}, \cite{ocker_pulsar_2023}).
In addition, the morphology of scintillation arcs sometimes varies over the course of weeks to months, as exemplified by the pulsar B1508+55 (\cite{Sprenger2022}).
Theoretical simulations often invoke the ``thin screen'' approximation, which takes into account the column density (=projected density) in the region of increased ISM fluctuations, see \cite{Cordes2006}.
Here, we compare the projected density approximation with simulations taking into account the three-dimensional plasma distribution (volume density).
We find that volume scattering leads to a richer and more asymmetric appearance of scintillation arcs, which is absent in the thin screen approximation.
The manuscript is organised as follows: In Sect.~\ref{sec:green} we review the basic formalism for radio wave propagation through a dielectric medium. 
Sects.~\ref{sec:volscatt}, \ref{sec:projscatt} describe the volume and projected scattering theories, which are applied in Sect.~\ref{sec:ism} to density models of the ISM.
Sect.~\ref{sec:secspec} discussed the resulting dynamic and secondary spectra and how they differ from the thin-screen approximation.
The SI includes details of the chosen potential and 150 additional examples of secondary spectra along with the three-dimensional plasma densities.

\section{Radio wave propagation in the interstellar space}\label{sec:green}

For the description of radio wave propagation through a dielectric medium with permittivity $\epsilon$, we introduce the three-dimensional potential 
\begin{equation}
V(\mathbf{r}')=(\epsilon(\mathbf{r}')-\epsilon_{\rm background})/(4\pi) \ne 0,
\end{equation}
which contains the electron number density $n_e$, which in turn determines the plasma frequency $\omega_p$ and the refractive index $n$ of the medium:
\begin{equation}
    \epsilon(\mathbf{r'}) = n^2(\mathbf{r'}) = 1-\frac{\omega_p^2(\mathbf{r'})}{\omega^2}, \quad \omega_p^2(\mathbf{r'})=\frac{n_e(\mathbf{r'}) e^2}{\epsilon_0 m_{e}}.
\end{equation}
For the study of pulsar scintillation arc formation and simplicity, we ignore the background electron density outside the ISM concentrations.
Maxwell's equation for the electric field $\mathbf{E}$ scattered by a dielectric becomes \cite[Eqs.~(50.5), (50.34)]{Schwinger1998}
\begin{equation}\label{eq:efield}
-\left(\nabla^2+k^2\right) \mathbf{E} =\frac{4 \pi}{c} \rmi k\left( \mathbf{J} +\frac{1}{k^2} \nabla ( \nabla \cdot \mathbf{J} )\right),\quad \text{with} \quad \frac{4 \pi}{c}\mathbf{J}=-i k (\epsilon-1) \mathbf{E} \quad \text{and} \quad \quad k=\frac{\omega}{c}=\frac{2\pi\nu}{c}.
\end{equation}
Neglecting changes of polarisation caused by 
$\nabla ( \nabla \cdot \mathbf{J} )$ in Eq.~(\ref{eq:efield}) leads to the stationary Helmholtz equation for an electric field component $E$
\begin{equation}\label{eq:helmholtz}
    \nabla^2 E+\frac{\omega^2}{c^2}\left(1-\frac{\omega_p^2(\mathbf{r})}{\omega^2}\right)E=0.
\end{equation}

\section{Volume scattering theory}\label{sec:volscatt}

As boundary conditions for the partial differential eq.~(\ref{eq:helmholtz}) we take an incoming plane wave $E_{\nu}(\mathbf{r},t)=E_0 \rme^{ikz-\omega t}$ for $z \rightarrow -\infty$, representing a distant pulsar with respect to the scattering region.
The dimensionless potential $V_{dl}(\mathbf{r})$ in the Helmholtz equation is given by
\begin{equation}\label{eq:vdl}
    V_{dl}(\mathbf{r})=\frac{\omega_p^2(\mathbf{r})}{\omega^2}=\frac{n_e(\mathbf{r}) e^2}{\epsilon_0 k^2 m_{e} c^2}.
\end{equation}
We solve the Helmholtz equation along a sequence of $N_z$ slabs across the dimensionless potential, similar to \cite{Fleck1976}:
\begin{equation}\label{eq:multisclice}
    E_{\nu,i}(x,y,z=N_z\Delta z) \approx \rme^{-\rmi k^2 V_{dl} \Delta z/(4k)}
    \left[ \prod_{j=1}^{N_z}  {\cal F}^{-1}
    \rme^{-\rmi p^2 \Delta z/(2k)}{\cal F}\rme^{-\rmi k^2 V_{dl}(x,y,j\Delta z)\; \Delta z/(2k)}
    \right]
    \rme^{+\rmi k^2 V_{dl}\Delta z/(4k) } E_{\nu,i}(x,y,z=0),
\end{equation}
with $p^2=-\partial_x^2-\partial_y^2+k^2$.
Here, ${\cal F}$ (${\cal F}^{-1}$) denotes the (inverse) two-dimensional Fourier transform along the $x$-$y$ plane.
Scintillation patterns are recorded by considering the time-dependent pulsar position at 
$\mathbf{r}_p(t)=(0,0,z_p)+\mathbf{v}_p t$, ($z_p\ll0$), and the time-dependent observer at $\mathbf{r}_o(t)=(0,0,z_o)+\mathbf{v}_o t$, ($z_o\gg0$).
For clarity we consider the case of a moving observer, while the pulsar and the dielectric remain stationary.
The case of an additionally moving pulsar is discussed in the SI and in the context of the Born approximation in \cite{Kramer2024b}.
An observer with velocity $\mathbf{v}_o=(v_{o,x},v_{o,y},0)$ records the position dependent dynamic intensities
${|E_\nu(\mathbf{r}_o+\mathbf{v}_o t)|}^2$ at times $[-\Delta_t / 2,\Delta_t / 2]$ for a band of frequencies $[\nu-\Delta_\nu / 2,\nu+\Delta_\nu / 2]$, which are Fourier transformed to obtain the secondary spectrum
\begin{equation}
{|H(f_t,f_\nu)|}^2={\left|  \int_{\nu_c-\Delta_\nu / 2}^{\nu_c+\Delta_\nu / 2} \rmd \nu\; \int_{-\Delta_t / 2}^{\Delta_t / 2} \rmd t\;\rme^{\rmi \nu 2\pi f_\nu+\rmi t 2\pi f_t}  {|E_\nu(\mathbf{r}_o+\mathbf{v}_o t)|}^2 \right|}^2.
\end{equation}

\section{Projected density Scattering theory}\label{sec:projscatt}

The dielectric is confined to a finite region along the line-of-sight (LOS) from the distant pulsar to the observer. 
If additionally the dielectric is located far away from the observer, Kirchhoff-Fresnel diffraction theory (KFD) is often invoked as a suitable approximation.
KFD is expressed in terms of the dispersion measure (DM), obtained by integrating the electron volume density along the propagation direction
\begin{equation}
    \text{DM}(x,y)=\int\rmd z\;n_e(\mathbf{r}),
\end{equation}
the KFD integral for the electric field intensity at the observer becomes
\begin{equation}\label{eq:KFD}
    {|E_\nu(\mathbf{r}_o)|}^2={\left|\frac{k}{2\pi\rmi z_o}\int\rmd x'\rmd y' \exp\left(\rmi k\frac{{(x_o-x')}^2+{(y_o-y')}^2}{2z_o}-\frac{\rmi}{2k} \frac{e^2}{\epsilon_0 m_e c^2} \text{DM}(x',y')\right)\right|}^2.
\end{equation}
KFD can be seen as the single-slab variant of the multi-slab method (\ref{eq:multisclice}) by projecting the potential along the $z$-axis to the $x$-$y$ plane intersecting the centre of mass of the electron density.
Additionally, the Fourier part is replaced in KFD by the free propagator in position space.
As discussed in Sect.~\ref{sec:secspec}, the projected density results differ in several aspects from the multi-slab method of the previous section.
KFD is invariant, i.e., \ produces an identical diffraction pattern, with scaling
\begin{equation}
    x\rightarrow x/L, \quad 
    y\rightarrow y/L, \quad
    z\rightarrow z/L^2, \quad
    k\rightarrow k \quad \text{(no change)}.
\end{equation}
This scaling works with the assumption that the scattering region only covers a small fraction of the total propagation length.
This implies that we can implement the numerical simulation on a grid with a more numerically convenient aspect-ratio compared to the actual geometry.
\cite{Arkadiew1913} performed experiments to confirm this relation and found agreement in his photographs with the analytical solution obtained by \cite{Lommel1886}.
Another common approach \citep{Rickett1984} is to introduce the wave number and screen-distance-dependent Fresnel scale $R_f=\sqrt{z/k}$ and dimensionless variables $\xi=x/R_f$, $\eta=y/R_f$.

\section{Pulsar signals from volume scattering by the ISM}\label{sec:ism}

To illustrate the impact and significance of the volume scattering across an extended but finite region as compared to the projected density scattering theory, we computed the secondary spectra in the presence of aligned and extended plasma distributions along the LOS.
The plasma density is restricted to a small region far away from the observer, and thus the scattering is, in principle, amendable to KFD.
Electron densities are expressed in terms of dimensionless potential (\ref{eq:vdl}).
Although the dimensionless potential does carry a dependency $1/k^2$, the term $k^2$ in Eq.~(\ref{eq:multisclice}) cancels this dependency and in the following we fix the electron density across variations of the $k$ vector.
For definiteness and reproducibility, the potential consists of the superposition of $25$ randomly directed cosine waves, which in the limit of a large number of modes lead to a Gaussian random potential (see SI for details).
The model is motivated by the effects of magnetic fields in the interstellar medium, which lead to elongated and compressed density distributions, as shown in simulations by \cite{Beattie2021,Beattie2025}. 
\begin{figure}
    \begin{center}
    \includegraphics[width=0.85\linewidth]{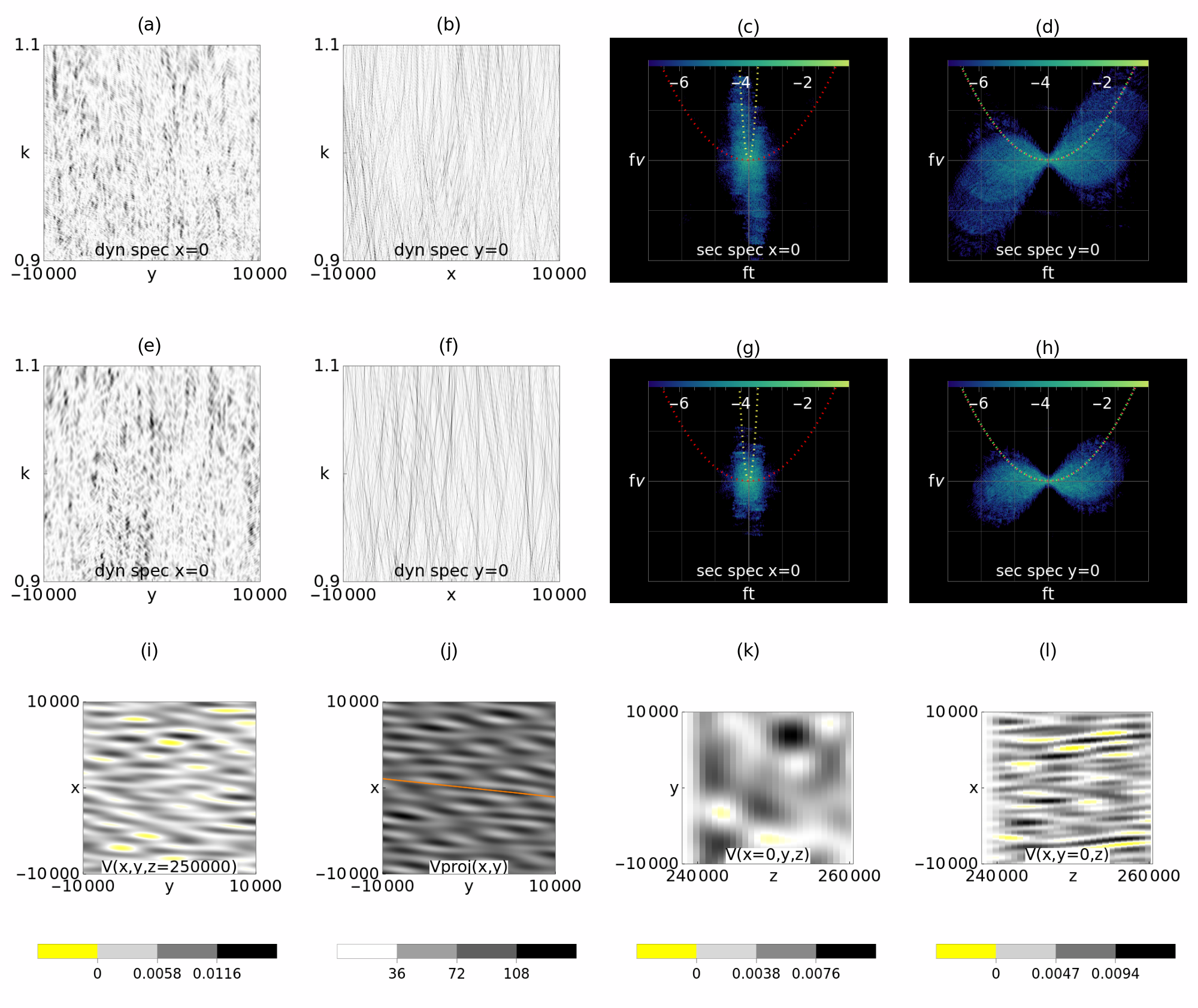}\\
    \includegraphics[width=0.85\linewidth]{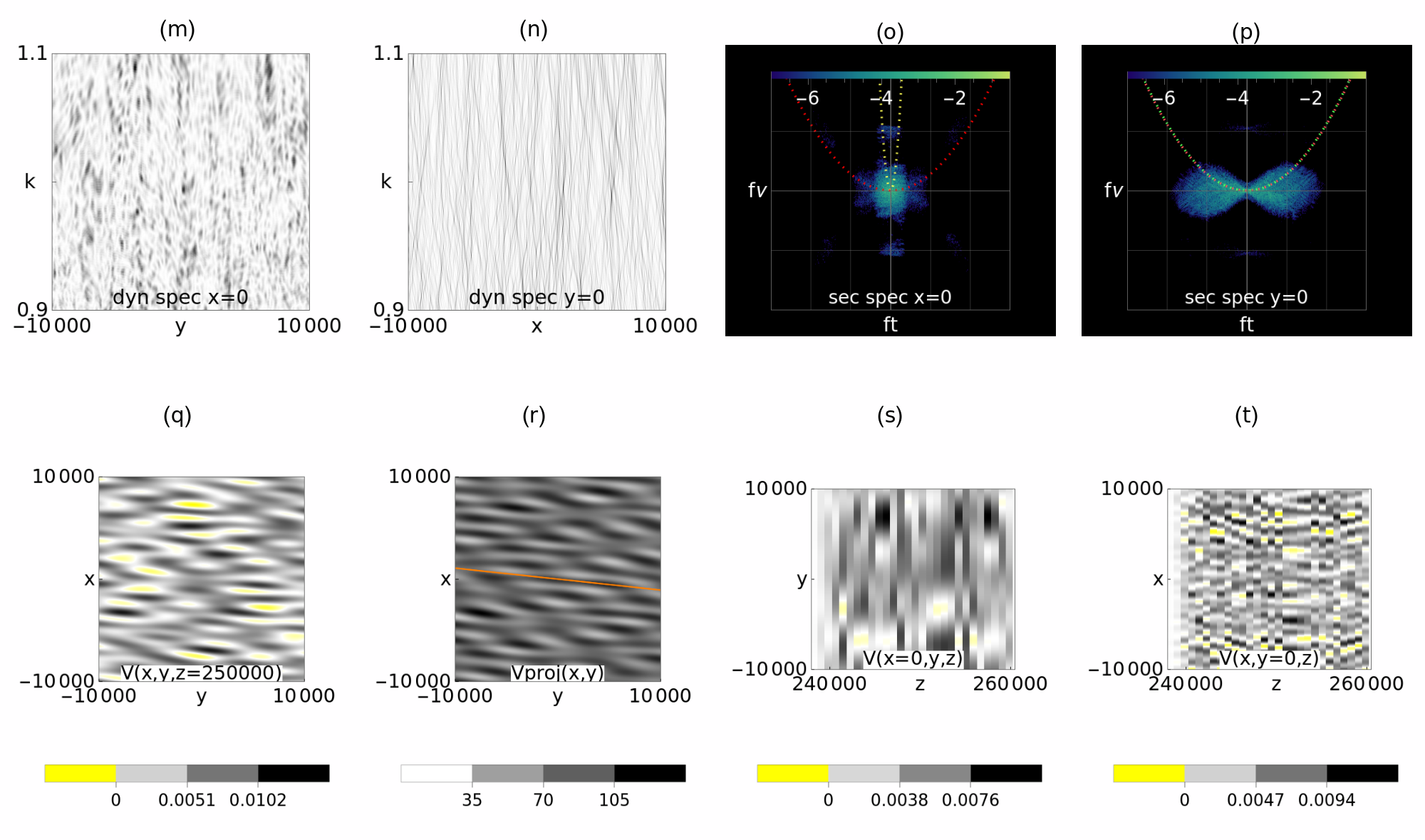}
    \end{center}
    \caption{
    Top row: volume scattering dynamic spectra for (a) $x_o=0, y_o=[-10000,10000], z_o=790000$ and 
    (b) $x_o=[-10000,10000], y_o=0, z_o=790000$ and corresponding secondary spectra, (c), (d).
    Second row: projected screen dynamic spectra (e) as in (a),  (f) as in (b) and corresponding secondary spectra (g), (h).
    Third row: (i), (k), (l) cuts through the scattering potential and projected (=column) density (j).
    Fourth and fifth row: same as above, but for a random permutation of the potential slabs along the $z$-axis (panels (s), (t)).
    The projected-density secondary-spectra (g),(h) are symmetric, while volume scattering leads to strong asymmetries of the arcs.
    The red parabola is the scintillation arc for scatterers parallel to the moving observer,
    the yellow and green parabola takes into account the angle of the dominant scattering structures of the projected density, indicated by the orange line in panel (j).
    }
    \label{fig:sheets2}
\end{figure}
Panels (i), (k) and (l) in Fig.~\ref{fig:sheets2} show cuts at $z=250000$, $x=0$ and $y=0$, respectively, through the volume electron density.
The projected electron density (column density along the $z$-axis) is shown in panel (j).
The orange line in panel (j) denotes the dominant orientation of the projected potential determined by the autocorrelation function of the projected density.\\
The potential is confined to a volume with extensions $[-20000:20000] \times [-20000:20000] \times [230000:270000]$ along the $x,y,z$ axes, the observer is located at $z_o=790000$.  
In the above parameters and in the figures, the length scale along the dimensions of $x$ and $y$ is reduced by $L=2000000$ and along $z$ by $L^2$.
In terms of the Fresnel scale, the transverse extension is about $b_x=26 R_f, b_y=26 R_f$.
This translates into an ISM model, where the fluctuating interstellar medium is LOS extended $2.6$~parsec along the LOS, with the observer located $71$~pc away from the centre of the scattering volume.
This corresponds to a scattering region restricted to $1/30$ of the distance from the scattering region to the observer.
Typical values of the dimensionless potential are on the order of $4/1000$, leading to $\text{DM}_\text{cloud}=7.3 \times 10^{-8}\;\text{parsec cm}^{-3}$ at $k=1$~m${}^{-1}$.
The tilted elongated structures in the electron density cause a more uniform areal electron density in projection, see panel (j).
We have investigated a total of 151 cases with differently rotated and stretched structures, with only the representative case \# 2 shown in the manuscript and the other in the SI.

\section{Dynamic and secondary spectra arising from volume vs.\ projected scattering}\label{sec:secspec}

Dynamic spectra are shown in Fig.~\ref{fig:sheets2}, panels (a,e,m) along the line $x_o=0, y_o=[-10000,10000], z_o=790000$ and in panels (b,f,n) along $x_o=[-10000,10000], y_o=0, z_o=790000$, corresponding to two perpendicular velocity vectors of the observer with $|\mathbf{v}_o| t \in [-10000,10000]$. 
In physical units, the range $[-10000,10000]$ corresponds to $[-0.133,0.133] \;\text{a.u}$.

For the frequency axis in the dynamic spectra $2000$ equidistant points are sampled between $k\pm d_k$, $k=1$~m$^{-1}$, $d_k=0.1$~m$^{-1}$.
The scintillation index for the dynamic spectra varies with frequency and is on the order of unity.
The corresponding secondary spectra are obtained by a two-dimensional Fourier transform of panels (a,b,e,f,m,n) and are shown in panels (c,d,g,h,o,p) with a logarithmic colour scale.
The Fourier transform is performed with the standard Tukey window function, see \cite{Bingham1967}.
The convergence of all results has been established by recomputing the panels with double and halved resolution.

The red parabola is calculated from SI, Eq.~(B4) with $\alpha=90^\circ$ $v_{o,x}=0$ for panels (c,g,o) or with the same result for $\alpha=0^\circ$ $v_{o,y}=0$ for panels (d,h,p), corresponding to a line of scatterers parallel to the velocity of the moving observer.
The yellow and green parabola are obtained from Eqs.~(B5,B6) by setting $\alpha$ to the angle between the velocity of the observer and the dominant direction in the projected density (orange line in panel (j)).\\
It is instructive to compare the volume scattering results (a,b,c,d) with the scattering arising from the projected density approximation (e,f,g,h).\\
In the projected density approximation, the asymmetries of the parabolic arms disappear.
The asymmetries in the arcs are caused by the random, but correlated, direction changes during the propagation of the electromagnetic wave caused by correlated neighbouring slabs through the electron volume density.
In contrast, the projected electron density splits an incoming electromagnetic wave in roughly equal parts around a region of increased projected density.
In the projected density, therefore, no significant asymmetries between the left and right arms of the scintillation arcs are expected.
This is confirmed for the other 150 cases shown in the SI.\\
The second visible difference between volume scattering and projected density is the shorter total extension of the arms.
This is caused by the diminished contrast in the projected density (panel (j)) compared to the volume density (panels (i), (j), (l)), where waves can meander around the plasma concentrations.\\
To investigate the impact of the correlation within the potential along the propagation direction, we construct a variant of the potential with randomly permuted density slabs through the volume density.
The projected density is invariant under the permutation of the slabs along the $z$-axis, but the correlation between consecutive slabs is reduced.
The secondary spectra resulting from a randomly permuted potential (see Fig.~\ref{fig:sheets2}, (m)-(t)) show a more symmetric appearance and resemble the projected density calculation, Fig.~\ref{fig:sheets2} (g),(h).
We conclude that the correlation length along the propagation direction plays an important role and the considered cases are not described by the diffusive process of random walk models of extended scattering volumes discussed by \cite{Cordes1986}.\\
Volume scattering across random media leads for certain parameter ranges to the formation of branched flow (see \cite{Heller2021} for a review), while a projected density causes the formation of only one generation of caustics, as done in the analysis of the corrugated sheet model proposed by \cite{Pen2014}.

\section{Conclusion}\label{sec:conclusion}

Volume scattering influences the morphology of parabolic scintillation arcs, including asymmetry, curvature, and total extension.
The curvature of the scintillation arcs is sometimes affected by the ISM structure and may not always serve as a reliable indicator of the distances between the pulsar, the ISM and the observer:
the angle $\alpha$ between the dominant direction of the projected density and the velocity (see SI Eq.~(B4)) fails to predict the curvature of the scintillation arc for the cases 78, 92, 99, 112, 122, 129 shown in the SI.

The correlated disorder across a finite volume range introduces directional asymmetries in the flux that are lost in the projected density models or in randomly permuted potentials.
Asymmetries in pulsar scintillation arcs have been reported for many pulsars, see \cite{Stinebring2022}.
The regime of branched flow occurs in weak but correlated scattering scenarios across a finite extension.
Only for an extended medium with a very large number of scattering events, the central limit theorem leads to a scattering behaviour amendable to stochastic theories put forward by \cite{Tatarski1961} and \cite{Salpeter1967}.
The three-dimensional structure of the ISM leads to a rich morphology of pulsar scintillations, see \cite{ocker_pulsar_2023}, which calls for a three-dimensional modelling to interpret the appearance and long-time variability of scintillation patterns.
An example of shifting the observer position into adjacent $x$-ranges (corresponding to different observations epochs) is shown in the SI, Fig.~C1, demonstrating the correlated movement of fine-structures in the secondary spectra along the main parabola over time.\\
In branched flow, a diffusive smearing of intensities is delayed by almost co-propagating bundles of ray paths.
Branched flow with resulting random arc asymmetries is a possible cause for transient changes on a month to year scale of secondary spectra found in pulsar data.

\section*{Acknowledgements}

This work benefited from the NVIDIA Academic Hardware Grant \textit{Simulating branched flow with tensor processing units}, PI Kramer (2022). 
Simulations were performed on the MUSICA supercomputer as part of the Austrian Scientific Computing (ASC) infrastructure at Johannes Kepler University Linz, Austria.
We acknowledge helpful discussions with E.J.~Heller, T.~Sprenger, D.~Stinebring and J.~Cordes.

\section*{Data availability}

The data underlying this article will be shared on reasonable request to the corresponding author.

\bibliographystyle{mnras}

% Don't change these lines
\bsp	% typesetting comment
\label{lastpage}
\end{document}